\documentclass[12pt]{article}
\usepackage{amsmath}
\usepackage{cite}
\usepackage{relsize}
\usepackage{amssymb}
\usepackage{graphics}
\usepackage{epsfig}
\usepackage{epstopdf}
\usepackage{verbatim}
\usepackage{color}
\usepackage{subcaption}
\usepackage{multirow}
\usepackage{textcomp}
\usepackage{float}
\usepackage[dvipsnames]{xcolor}
\usepackage{mathtools}
\usepackage[toc,page]{appendix}
\usepackage{enumitem}

\begin{document}

\title{\textbf{Scalar Dark Matter in $A_4$ based texture one-zero neutrino mass model within Inverse Seesaw Mechanism}}

\author {\small{Rishu Verma\thanks{ rishuvrm274@gmail.com}, Monal Kashav\thanks{ monalkashav@gmail.com}, Surender Verma\thanks{ s\_7verma@yahoo.co.in} and B. C. Chauhan\thanks{ bcawake@hpcu.ac.in}}}

\date{\textit{Department of Physics and Astronomical Science,\\Central University of Himachal Pradesh, Dharamshala 176215, INDIA.}}

\maketitle

\begin{abstract}
In this paper, we present a model based on $A_4$ discrete flavor symmetry implementing inverse and type-II seesaw mechanisms to have LHC accessible TeV scale right-handed neutrino mass and texture one-zero in the resulting Majorana neutrino mass matrix, respectively. We investigate neutrino and dark matter sectors of the model. Non-Abelian discrete $A_{4}$ symmetry spontaneously breaks into $Z_{2}$ subgroup and hence provide stable dark matter candidate. To constrain the Yukawa Lagrangian of our model, we imposed $Z'_2$, $Z_3$ and $Z_4$ cyclic symmetries in addition to the $A_4$ flavor symmetry. In this work we used the recently updated data on cosmological parameters from PLANCK 2018.  For the dark matter candidate mass around 45 GeV-55 GeV, we obtain the mediator particle mass(right-handed neutrinos) ranging from 138 GeV to 155 GeV. The Yukawa couplings is found to be in the range 0.995-1 to have observed relic abundance of dark matter. We, further, obtain inverse ($X\equiv\frac{F^2n}{z^2}$) and type-II ($X^{'}\equiv f_1 v_{\Delta_{1}}$) seesaw contributions to $0\nu\beta\beta$ decay amplitude $|M_{ee}|$, while model being consistent with low energy experimental constraints. In particular, we emphasize that type-II seesaw contribution to $|M_{ee}|$ is large as compared to inverse seesaw contribution for normally ordered(NO) neutrino masses.
\end{abstract}

\section{Introduction}
Standard Model of particle physics is a low energy effective theory which has been astonishingly successful in explaining the dynamics of fundamental particles and their interactions. The discovery of Higgs Boson in 2012 at CERN LHC has strengthen our belief in this incredible theory. Despite its tremendous success there still remain unanswered questions such as origin of neutrino mass, dark matter and matter-antimatter asymmetry, to name a few. Neutrino oscillation experiments have been very instrumental in our quest to understand Standard Model(SM) predictions for the leptonic sector. They have shown at high level of statistical significance that neutrinos have non-zero but tiny mass, and flavor and mass eigenstates mix giving rise to quantum mechanical phenomena of neutrino oscillations.

\noindent On the contrary, within the SM, neutrinos are massless because the Higgs field cannot couple to the neutrinos due to the absence of right-handed(RH) neutrinos. The extension of SM with RH neutrino, however, require unnatural fine tuning of the Yukawa couplings to generate sub-eV neutrino masses. Dimension five Weinberg operator can generate the tiny Majorana mass for neutrinos with the SM Higgs field. In fact, there exist several beyond the Standard Model(BSM) scenarios, e.g. seesaw mechanisms, which may explain the origin of such dimension five operator and can account for dynamical origin of tiny Majorana neutrino masses by appropriately extending the field content of the SM. For example in type-I, type-II and type-III seesaw RH neutrinos, scalar triplet(s) and fermion triplet(s) are introduced to the particle content of the SM, respectively\cite{s1,s2,s3,s4,s5,s6}.

\noindent Seesaw mechanisms provide the most elegant and natural explanation of the smallness of neutrino masses. The fundamental basis of seesaw mechanism is the existence of lepton number violation at some high energy scale. In order to have neutrino mass at sub-eV, the new physics scale must be of the order of GUT scale, 10$^{16}$ GeV. Therefore, the seesaw mechanisms explain the tiny non-zero neutrino masses, but they introduce new physics scale, which is beyond the reach of current and near future accelerator experiments.

\noindent On the other hand, there are several astonishing astrophysical observations such as (i) galaxy cluster investigation\cite{gc1}, (ii) rotation curves of galaxy\cite{rc1}, (iii) recent observations of bullet cluster\cite{bc1}, and (iv) the latest cosmological data from Planck collaboration\cite{planck2018}, which have proven the existence of non-luminous and non-baryonic anatomy of matter known as ``Dark Matter"(DM). Apart from the astrophysical environments, it is very difficult to probe the existence of DM in terrestrial laboratory. Alternatively, one can accredit a weak interaction property to the DM through which it get thermalized in the early Universe, which also can be examined at the terrestrial laboratories. According to the Planck data the current DM abundance is\cite{planck2018} 
\begin{center}
$\Omega_{DM}h^2 = 0.120\pm0.001$.
\end{center}

\noindent Apart from relic abundance, the particle nature of DM is still unknown. Within the SM of particle physics, there is no suitable candidate for DM, as it should be stable on the cosmological time scale. The uncertainty in the nature of DM and possible mechanisms of neutrino mass generation have opened up a window to explore new models in a cohesive way. 
 Out of several neutrino mass models, the texture zero models are very interesting due to their rich phenomenology and high predictability\cite{t1,t2,t3,t4,t5,t6,t7,t8,t9,t10}. In fact, texture zero models have also been successfully investigated to generate observed matter-antimatter asymmetry in different seesaw settings\cite{t11} and scotogenic scenarios\cite{sctt}. The DM phenomenology within the framework of texture zeros using type-I seesaw and scotogenesis have been studied in Refs. \cite{type1} and \cite{sct}, respectively.

 \noindent Another possibility amongst the phenomenological approaches is the existence of scaling structure in the neutrino mass matrix wherein third column is scaled with respect to the second column by some model dependent parameter(s)\cite{rnm,svrma}. The scaling ansatz has been studied in Ref.\cite{mnd} where neutrino mass is generated using inverse and type-II seesaw frameworks. In Ref.\cite{mnd}, the implementation of inverse seesaw resulted in scaled neutrino mass matrix which predicts vanishing lightest mass eigenvalue(inverted neutrino mass ordering) and vanishing reactor mixing angle($\theta_{13}=0$). Subsequently, within an effective field theory approach, in order to have non-zero reactor mixing angle($\theta_{13}\neq 0$) type-II seesaw perturbation has been incorporated in the Lagrangian. We, in this work, confine to tree level dimension-4 and focus on the possible realization of texture one-zero ansatz which can be embedded in more general framework of grand unified theories wherein quarks and leptons belong to the same multiplet. For example, in Refs.\cite{fuku, svrma1}, texture zero(s) in fermion mass matrices have been investigated under SO(10) environment. 
 
\noindent In this work, we investigate a well-motivated possibility for simultaneous explanation of DM and non-zero neutrino mass using $A_4$ non-Abelian discrete symmetry within the framework of inverse seesaw(ISS)\cite{iss1,iss2,iss3} wherein small neutrino masses emanate from new physics at TeV scale, which is within the reach of accelerator experiments. The stability of DM is assured by $Z_2$ symmetry. Type-II seesaw has been implemented to have one-zero in the effective neutrino mass matrix. Within ISS mechanism, neutrino masses are generated assuming three right-handed neutrinos $N_{T}$ and three additional SM singlet neutral fermions $S_{T}$($T=1,2,3$). The fermionic singlets ($N_{4,5}$ and $S_{4,5}$) are assumed to have Yukawa couplings with the scalar fields $H,\phi,\phi_R$ and $\phi_{S}$, which after spontaneous symmetry breaking provide diagonal Majorana mass matrix($\mu$). The scalar triplets $\Delta_1$ and $\Delta_2$ are incorporated, so that $M_{\nu}$ contain one vanishing element after type-II seesaw implementation. Along with DM abundance, we have also obtained prediction of the model for effective Majorana mass ($|M_{ee}|$) appearing in neutrinoless double beta($0\nu\beta\beta$) decay.     

\noindent The paper is structured as follows. In section {\ref{sec:2}}, we have discussed the inverse seesaw mechanism based on $G_f\equiv A_4\times Z_{2}^{'}\times Z_3\times Z_4$ symmetry group and resulting neutrino mass matrices. Section  {\ref{sec:3}} is devoted to the investigation of relic density of the DM. In section  {\ref{sec:4}}, the prediction of the model for neutrinoless double beta decay is discussed. Finally, conclusions are summarized in section  {\ref{sec:5}}.

\section{The Model}\label{sec:2}
In order to explain the smallness of neutrino mass, different versions of the seesaw mechanism play an important role. As discussed above the ISS mechanism is a viable scenario to get the mass of right-handed neutrino near the TeV scale. This scale is much below the scale, which we get from the canonical seesaw. As a requirement of ISS, the fermion sector is extended by three right-handed neutrinos $N_{i} (i = 1,2,3)$ and three extra singlet fermions $S_{j}(j=1,2,3)$. Within ISS mechanism the mass Lagrangian is written as\\
\begin{equation}
L = -\bar{\nu}_{\alpha L}m_{D}N_{i} - \bar{S_{j}}mN_{i}-\frac{1}{2}\bar{S_{j}}\mu S^C_{k} + h.c. \label{eq:1}
\end{equation}
where $m_D$, $m$ and $\mu$ are the 3$\times$3 complex mass matrices and $\alpha = (e,\mu,\tau)$, $k=(1,2,3)$. Here $m$ represents lepton number conserving interaction between neutral fermions and right-handed neutrinos and $\mu$ gives the Majorana mass terms for neutral fermions. Assuming lepton number as approximate symmetry, the Majorana mass term for right-handed neutrino is vanishing. However, it is mildly violated through singlet fermions $S$ having small mass $\mu$ in consonance with 't Hooft's naturalness criterion\cite{hooft}. The lepton number symmetry can be restored as $\mu\rightarrow 0$. Consequent to spontaneous symmetry breaking (SSB), the Lagrangian in Eqn.(\ref{eq:1}) leads to 9$\times$9 neutrino mass matrix\\
\begin{equation}
 M_{\nu} = \begin{pmatrix}
0 & m_{D} & 0\\
m_D^T & 0 & m\\
0 & m^T & \mu \\
\end{pmatrix}, \\
\label{eq:2}
\end{equation}
in the basis ($\nu_L,N,S$). We can obtain standard model neutrinos at sub-eV scale from $m{_D}$ at electroweak scale, $\mu$ at keV scale and $m$ at TeV scale as explained in \cite{iss3,33}. Thus, if we consider the order $\mu << m_D << m$, then after the block diagonalization of above matrix, the 3$\times$3 effective neutrino
mass matrix is obtained as\\
\begin{equation}
 m_{\nu} = m_{D} (m{^T})^{-1}\mu m^{-1} m_{D}^{T}.\\
 \label{eq:3}
\end{equation}

\noindent It is clear from Eqn.(\ref{eq:3}) that there is a double suppression by mass term associated with $m$, which results in the scale that is much below to the one obtained by canonical seesaw. The essence of inverse seesaw mechanism lies in the fact that we can bring down the mass of right-handed neutrinos to TeV scale by assuming that $\mu$ should be at keV scale \cite{34,35,36}.

\noindent The symmetry group $A_4$ has played an important role in understanding particle physics \cite{37,38,39,40,41}. $A_4$ is a non-Abelian discrete symmetry group of even permutations of four objects.
Order of this group is 12. All the 12 elements are generated from two elements, S and T which satisfies: $S^2$ = $T^3$ = $(ST)^3$. It is a symmetry group of regular tetrahedron. It has four conjugacy classes, therefore, four irreducible representations: 1, 1$'$, 1$''$ and 3. The multiplication rules of irreducible representations in T basis are \cite{41,42}: {\bf1}$'\otimes${\bf1}$'$={\bf1}$''$,
{\bf1}$''\otimes${\bf1}$''$={\bf1}$'$, {\bf1}$'\otimes${\bf1}$''$={\bf1}, %{\bf3}$\otimes${\bf3}={\bf1}$\oplus${\bf1}$'\oplus${\bf1}$''\oplus${\bf3}$_s\oplus$
%{\bf1}$''\otimes${\bf1}$''$={\bf1}$'$,
%{\bf1}$'\otimes${\bf1}$''$={\bf1},
{\bf3}$\otimes${\bf3}={\bf1}$\oplus${\bf1}$'\oplus${\bf1}$''\oplus${\bf3}$_s\oplus${\bf3}$_a$ where,
\begin{eqnarray}
\nonumber
&&\left(\bf{3}\otimes\bf{3}\right)_{\bf{1}} =a_1b_1+a_2b_2+a_3b_3,\\ \nonumber
&&\left(\bf{3}\otimes\bf{3}\right)_{\bf{1'}}=a_1b_1+\omega a_2b_2+\omega^2 a_3b_3,\\ \nonumber
&&\left(\bf{3}\otimes\bf{3}\right)_{\bf{1''}}=a_1b_1+\omega^2 a_2b_2+\omega a_3b_3,\\ \nonumber
&&\left(\bf{3}\otimes\bf{3}\right)_{\bf{3_s}}=\left(a_2b_3+b_2a_3,a_3b_1+a_1b_3,a_1b_2+a_2b_1\right),\\ \nonumber
&&\left(\bf{3}\otimes\bf{3}\right)_{\bf{3_a}}=\left(a_2b_3-b_2a_3,a_3b_1-a_1b_3,a_1b_2-a_2b_1\right). \nonumber
\end{eqnarray}

\noindent Here $a_{i}$ and $ b_{i}$ (i = 1,2,3)  are the basis vectors of the two triplets and $\omega$ = $e^{\frac{2\pi i}{3}}$.

\noindent In the model we have taken five right-handed neutrinos, three of which $N_T = (N_1,N_2,N_3)$ are transforming as triplet under $A_4$ and rest of the two i.e., $N_4$ and $N_5$ are transforming as singlets $1, 1'$, respectively. Singlet fermions $S_T = (S_1,S_2,S_3)$ and $(S_4, S_5)$ transforming as triplet and singlets $(1, 1')$ under $A_4$, respectively, are also introduced. The standard model Higgs doublet $H$ and three additional Higgs doublets $\eta_i$ transform as singlet and triplet $\eta$ under $A_4$, respectively.  In addition, we have extended the scalar sector with three $SU(2)_L$ singlet scalar fields i.e. $\phi$, $\phi_R$ and $\phi_S$. After spontaneous symmetry breaking (SSB), the vacuum expectation values ($vev$) acquired by ($H$, $\eta$) and ($\phi$, $\phi_R$, $\phi_S$) give $m_D$ and ($m$, $\mu$), respectively, with minimal number of parameters. In order to have possible lepton number violation via $S_i (i=T,4,5)$ only, we distinguish the Yukawa interactions of $N_i$ and $S_i$ through $Z_2^{'}$ symmetry. Also, using $Z_3$ symmetry, all possible $N_iN_j(i,j=T,4)$ Majorana terms are inhibited. There can be possibility of Yukawa interactions like $N_TN_T\phi_{S}^{*}$, $N_4N_4\phi_{S}^{*}$ which are suppressed by $Z_4$ symmetry in the model. The fermionic and scalar field content along with respective charge assignments are shown in Table \ref{table1} and Table \ref{table2}, respectively.

\begin{center}
\begin{table}[t]
\centering
\begin{tabular}{ccccccccccccc}
 Symmetry & $\bar{L}_{e}$ & $\bar{L}_{\mu}$ & $\bar{L}_{\tau}$ & $e_{R}$ & $\mu_{R}$ & $\tau_{R}$ & $N_{T}$ & $N_{4}$ & $N_{5}$ & $S_{T}$ & $S_{4}$ & $S_{5}$\\
 \hline
$SU(2)_L$    &     2  & 2 & 2 & 1  & 1 & 1   & 1   &   1     &    1        &   1        &    1   & 1 \\
\hline
$A_{4}$ & 1   & 1$'$ & 1$''$  &      1    &   1$''$ &  1$'$ & 3 & 1 &  1$'$  & 3 & 1 &  1$'$ \\
\hline
$Z'_{2}$     & 1   & 1 & 1
  &      1    &   1  &  1 & 1 & 1 &  1  & -1 & -1&  -1 \\
  \hline
$Z_{3}$ & $\omega^2$  & $\omega^2$ & $\omega^2$  &    1    & 1  &  1  & $\omega^2$ & $\omega^2$ &  $\omega^2$  & $\omega$ & $\omega$ &  $\omega $\\
\hline
$Z_{4}$ & 1  & 1 & 1  &     1    & 1 &  1  & 1& 1 &  1 & $i$ & $i$ & $i$\\
\hline
\end{tabular}
 \caption{Fermion field content and respective charge assignments used in the model.}
 \label{table1}
\end{table}
\end{center}
\begin{center}
\begin{table}[h]
\centering
\begin{tabular}{cccccccc}
  Symmetry & H & $\eta$ & $\Phi$ & $\Phi_{R}$ & $\Phi_{S}$ & $\Delta_{1}$ & $\Delta_{2}$ \\ 
 \hline
  $SU(2)_L$    &    2 & 2    & 1 & 1 & 1 & 3 & 3               \\
\hline
$A_{4}$ & 1       &   3 &   3  & 1    &    1     & 1 & 1$'$ \\
 \hline
 $Z'_{2}$     & 1       &   1 &   -1  & -1    &    1     & 1 & 1 \\
 \hline
   $Z_{3}$ &   $\omega$       &   $\omega$ &   1  & 1    &    $\omega $     & $\omega $ & $\omega $ \\
 \hline
  $Z_{4}$ &   1       &   1 &   $-i$  & $-i$    &   - 1      & 1 & 1 \\
 \hline
 \end{tabular}
 \caption{Scalar field content and respective charge assignments used in the model.}
\label{table2}
\end{table}
\end{center}

\noindent The leading Yukawa Lagrangian is
\begin{eqnarray}
\nonumber
\mathcal{L^I} =&&y_{e}\Bar{L}_{e}e_{R}H + y_{\mu}\Bar{L}_{\mu}e_{\mu}H + y_{\tau}\Bar{L}{_\tau}e_{\tau}H + y_{1}^{\nu}\Bar{L}_{e}[N_{T} \Tilde{\eta}]_{1} + y_{2}^{\nu}\Bar{L}_{\mu}[N_{T} \Tilde{\eta}]_{1''}+\\
 \nonumber
&&y_{3}^{\nu}\Bar{L}_{\tau}[N_{T} \Tilde{\eta}]_{1'} + y_{4}^{\nu}\Bar{L}_{e}N_{4}\Tilde{H} + y_{5}^{\nu}\Bar{L}_{\tau}N_{5}\Tilde{H} + y_{R}^{1}[N_{T} S_{T}]_{1} \phi_{R}+\\
\nonumber
 &&y_{R}^{0}[N_{4} S_{4}]_{1} \phi_{R} + y_{\phi}^{1}[N_{T} \phi]_{1} S_{4} + y_{\phi}^{2}[N_{T} \phi]_{1''} S_{5} + y_{\phi}^{3}N_{4} [\phi S_{T}]_{1} + y_{\phi}^{4}N_{5} [\phi S_{T}]_{1''} + \\
 \nonumber
 && y_{\phi}^{5}[N_{T} \phi]_{3} S_{T} + y_{s}^1 S_{T}S_{T} \phi_{S} + y_{s}^2 S_{4}S_{4} \phi_{S}  + h.c.,\\
 \label{eq:4}
\end{eqnarray}

\noindent where $\Tilde{H}=i\tau_3H$, $\Tilde{\eta}=i\tau_3\eta$ and $y_q(q = e,\mu,\tau)$, $y_{i}^{\nu} (i = 1,2,3,4,5)$, $y_{R}^{j}(j=0,1)$, $y_{\phi}^{k}(k=1,2,3,4,5)$, $y_{s}^p(p=1,2)$ are Yukawa coupling constants.
We have chosen the following vacuum alignments 
\begin{center}
$\langle \eta \rangle \sim v_{\eta} (1,0,0)$,
$\langle \phi \rangle \sim  v_{\phi}(1,0,0)$,
$\langle H \rangle = v_{h},\langle \phi_{S} \rangle = v_{S}, \langle \phi_{R} \rangle = v_{R}$.
\end{center}

 \noindent The symmetry is broken down to $Z_2$ subgroup i.e. $G_f\equiv A_4\times Z_{2}^{'}\times Z_3\times Z_4\rightarrow Z_2$  by the $vev$ $\langle \eta \rangle \sim v_{\eta} (1,0,0)$ \cite{ddm}. Since the $vev$ alignment  (1,0,0) remains invariant under the A$_4$ generator S = $Diag(1,-1,-1)$, the residual Z$_2$ symmetry is
\begin{center}
N$_2\rightarrow$ -N$_2$,      N$_3\rightarrow$ -N$_3$, 
S$_2\rightarrow$ -S$_2$,      S$_3\rightarrow$ -S$_3$,\\
$\eta_2\rightarrow -\eta_2$,  $\eta_3\rightarrow -\eta_3$,
$\phi_2\rightarrow -\phi_2$, $\phi_3\rightarrow -\phi_3$.
\end{center}
It is to be noted that only residual symmetry $Z_2$ is responsible for lightest dark matter candidate stability while $Z_2^{'}$ is employed to restricts the unwanted Yukawa couplings. $Z_2^{'}$ does not play any role in dark matter stabilization as the DM candidate is $Z_2^{'}$ even. From Eqn.(\ref{eq:4}), it is evident that lepton conserving interactions of $N_T$ and $S_T$ take place through the $A_4$ triplet $\phi$ introduced in model setup. Since inverse seesaw formula in Eqn.(\ref{eq:3}) assumes hierarchy of mass scale $\mu << m_D << m$, which implies that lepton conserving interaction of right-handed neutrinos $N_T$ and neutral fermion $S_T$ takes place at high scale.   Also, Dirac mass term emanates from the $A_4$ triplet of SU(2)$_L$ doublet Higgs field $\eta$ having smaller mass as compared to interactions involving  $N_T$,  $S_T$. So, lightest $Z_2$ stabilized candidate will be $\eta_{2,3}$. It couples only with the right-handed neutrinos and not with the charged leptons. As a consequence, we have obtained a diagonal charged lepton mass matrix as\begin{equation}
m_{l}= Diag(y_{e},y_{\mu},y_{\tau})v_{h}.\\
\end{equation}   
 
\noindent The other mass matrices which we have obtained are shown as below :\\
\begin{equation}
m_{D} =\begin{pmatrix}
A & 0 & 0 & F & 0\\
B & 0 & 0 & 0 & 0\\
C & 0 & 0 & 0 & H\\
 \end{pmatrix}, \mu = \begin{pmatrix}
y & 0 & 0 & 0 & 0\\
0 & y & 0 & 0 & 0\\
0 & 0 & y & 0 & 0\\
0 & 0 & 0 & n & 0\\
0 & 0 & 0 & 0 & 0\\

\end{pmatrix} ,
m = \begin{pmatrix}
x & 0 & 0 & l & v\\
0 & x & h & 0 & 0\\
0 & h & x & 0 & 0\\
l & 0 & 0 & z & 0\\
v & 0 & 0 & 0 & 0\\
\end{pmatrix}, \\
\end{equation}

\noindent where $A=y_{1}^{\nu}v_{\eta}, B = y_{2}^{\nu}v_{\eta}, C=y_{3}^{\nu}v_{\eta}, F = y_{4}^{\nu}v_{h}, H = y_{5}^{\nu}v_{h}, $
$y = y_{S}^{1}v_{S}, n = y_{S}^{2}v_{S},$
$x= y_{R}^{1}v_{R}, z= y_{R}^{0}v_{R},l = y_{\phi}^{1}v_{\phi}+y_{\phi}^{3}v_{\phi}, v =  y_{\phi}^{2}v_{\phi}+y_{\phi}^{4}v_{\phi}, h =  y_{\phi}^{3}v_{\phi}$. Within ISS mechanism, the above matrices lead to the light neutrino mass matrix as follow\\
\begin{equation}
 m_{\nu_{I}} = \begin{pmatrix}
X & 0 & \Delta \\
0 & 0 & 0 \\
\Delta & 0 & \Delta^{''} \\
\end{pmatrix} ,\\
\end{equation}

\noindent where $X = \frac{F^2n}{z^2}$ , $\Delta = -\frac{FHln}{vz^2}$ and $\Delta^{''} = \frac{H^2(l^2n+yz^2)}{v^2z^2}$.

\noindent When we choose a flavor basis in which we are obtaining a diagonal charged lepton mass matrix, only those neutrino mass matrix are allowed where we have at most  two zeros. These neutrino mass matrices are consistent with neutrino oscillation results \cite{texture01}. Since we are getting three zeros in our neutrino mass matrix, we have introduced type-II seesaw to reduce the number of zeros in the neutrino mass matrix. The type-II seesaw contribution to the Lagrangian is given as 
\begin{eqnarray}
\nonumber
 \mathcal{L^{II}} = &&f_{1}(L_{e}L_{e}+L_{\mu}{L}_{\tau}+L_{\tau}{L}_{\mu})\Delta_{1} +\\
 \nonumber
 &&f_{2}(L_{e}{L}_{\tau}+{L}_{\tau}{L}_{e}+{L}_{\mu}{L}_{\mu})\Delta_{2} + h.c.,\\
 \label{eq:8}
 \end{eqnarray}
 where, $f_1$ and $f_2$ are coupling constants. Therefore, the ISS + type-II seesaw Lagrangian for our model is given as
 \begin{eqnarray}
 \nonumber
 \mathcal{L} =&&y_{e}\Bar{L}{_e}e_{R}H + y_{\mu}\Bar{L}{_\mu}\mu_{R}H + y_{\tau}\Bar{L}{_\tau}\tau_{R}H + y_{1}^{\nu}\Bar{L}_{e}N_{T} \Tilde{\eta} + y_{2}^{\nu}\Bar{L}_{\mu}N_{T}\Tilde{\eta} +y_{3}^{\nu}\Bar{L}_{\tau}N_{T} \Tilde{\eta} +\\ 
 \nonumber
&&y_{4}^{\nu}\Bar{L}_{e}N_{4}\Tilde{H} + y_{5}^{\nu}\Bar{L}_{\tau}N_{5}\Tilde{H} +y_{R}^{0}N_{4} S_{4} \phi_{R} + y_{R}^{1}N_{T} S_{T} \phi_{R} +  y_{\phi}^{1}N_{T} \phi S_{4} +\\
\nonumber
 && y_{\phi}^{2}N_{T} \phi S_{5} + y_{\phi}^{3}N_{4}\phi S_{T} + y_{\phi}^{4}N_{5} \phi S_{T} + 
  y_{\phi}^{5}N_{T} \phi S_{T} + y_{s}^1 S_{T}S_{T} \phi_{S} + y_{s}^2 S_{4}S_{4} \phi_{S} +\\
  \nonumber
&&f_{1}(L_{e}L_{e}+L_{\mu}{L}_{\tau}+L_{\tau}{L}_{\mu})\Delta_{1} +f_{2}(L_{e}{L}_{\tau}+{L}_{\tau}{L}_{e}+{L}_{\mu}{L}_{\mu})\Delta_{2} + h.c.\\
\label{eq:9}
\end{eqnarray}

\noindent The SU(2) triplets $\Delta_{1}$ and $\Delta_{2}$ are transforming as singlets 1 and 1$'$, respectively. The vacuum expectation values  $ \langle \Delta_{1} \rangle = v_{\Delta_{1}} ,  \langle \Delta_{2} \rangle = v_{\Delta_{2}}  $ gives
\begin{equation}
m_{{\nu}_{II}} = \begin{pmatrix}
X^{'} & 0 & \Delta^{'} \\
0 & \Delta^{'} & X^{'} \\
\Delta^{'} & X^{'} & 0 \end{pmatrix},
\end{equation}

\noindent where, $X^{'} = f_1 v_{\Delta_{1}}, \Delta^{'}=f_2 v_{\Delta_{2}}. $

\noindent Finally, the neutrino mass matrix is $M_{\nu}=m_{\nu_{I}}+m_{\nu_{II}}$ and can, explicitly, be written as\\\\
  \begin{equation}
  M_{\nu}= 
  \begin{pmatrix}
    X+X^{'}& 0 & \Delta + \Delta^{'}  \\
     0 & \Delta^{'}   & X^{'}  \\
    \Delta + \Delta^{'} & X^{'} & \Delta^{''}\
   \end{pmatrix}.
   \label{eq:11}
 \end{equation}
 
 \noindent In literature, there are several techniques used to reduce the parameters of neutrino mass matrix, and texture zeros is one of them \cite{texture01,texture02,texture03,texture04,texture05,text006,text07,text08,text09,text010,text011,text012,text013,text014,text015,text016,text017,text018,text019,text020}. Interestingly, the mass matrix in Eqn.(11) corresponds to texture one-zero neutrino mass model. The phenomenological implications of these class of models have been extensively studied in the literature \cite{text015,text016,text017,text018}. Rather, in the present work we study the prediction of the current setup for  beyond neutrino sector observable like DM and $0\nu\beta\beta$ decay discussed below. 
 
\section{Relic Density of Dark Matter}\label{sec:3}
In the early universe, the particles were in thermal equilibrium i.e. the processes in which the lighter particles combine to form heavy particles and vice-versa happened at same rate. At some point of time, the conditions required for thermal equilibrium were contravened because the density of some particle species became too low. These particles are stated as ``freeze-out" and they have a constant density which is known as relic density, because the abundance of particle remains same. In the process, if any particle $\chi$ was in thermal equilibrium, then its relic abundance can be obtained by using Boltzmann equation\cite{relic,relic1,relic2,relic3}
\begin{equation}
\frac{dn_{\chi}}{dt} +3\mathcal{H}n_{\chi}= -<\sigma v>(n_{\chi}^2 - (n_{\chi}^{eqb})^2),
\label{eq:12}
\end{equation}

\noindent where $\mathcal{H}$ is the Hubble constant and $ n_{\chi}$ is the number density of the DM particle $ \chi$. Here, $n_{\chi}^{eqb}$ is the number density of particle $ \chi$ when it was in thermal equilibrium. However, $<\sigma v>$ is the thermally averaged annihilation cross-section of the DM particle. For a DM particle with electroweak scale mass, the solution of above Eqn.(\ref{eq:12}) gives \cite{approx}
\begin{equation}
\Omega_{\chi}h^2 = \frac{3 \times 10^{-27} cm^3 s^{-1 }}{<\sigma v>},\\
\label{eq:13}
\end{equation}

\noindent where $\Omega_{\chi}h^2$ gives the relic density of DM particle.

\begin{figure}[t]
\begin{center}
\includegraphics[scale=.8]{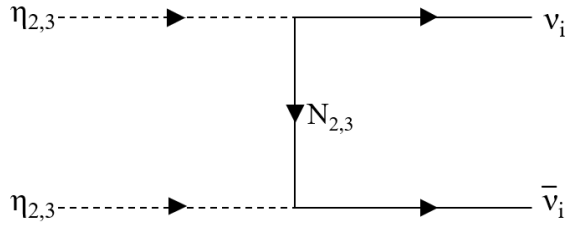}
\end{center}
\caption{Scattering of DM particle $\eta_{2,3}$.}
\label{fig1}
\end{figure}

\begin{figure}[t]
	\begin{center}
			{\epsfig{file=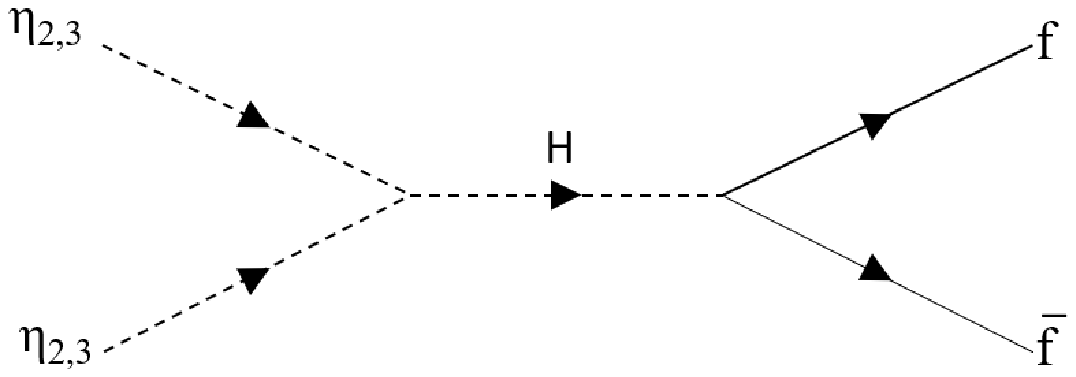,height=3.2cm,width=7.0cm}
				\epsfig{file=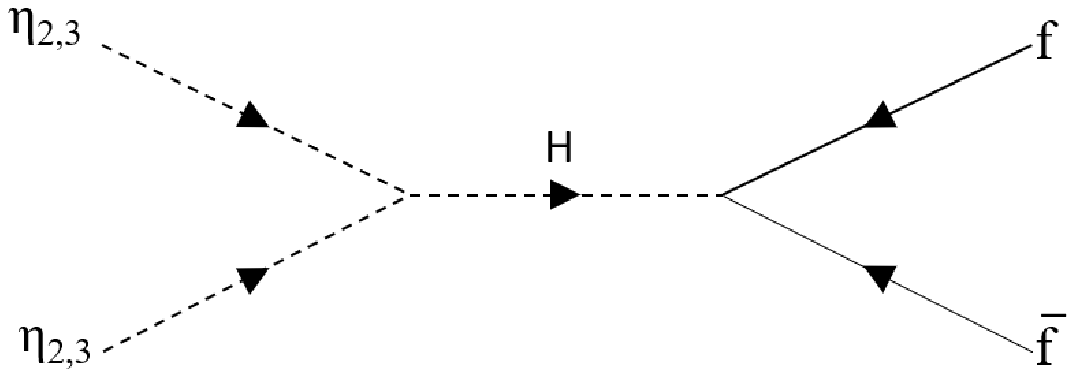,height=3.0cm,width=7.0cm}}
		\end{center}
\caption{\label{fig2}Self annihilation of DM particle $\eta_{2,3}$ \cite{fig}.}
\end{figure}

\noindent From the Lagrangian in Eqn.(\ref{eq:9}), the interaction of dark matter particle with right-handed neutrinos is as shown in Fig.\ref{fig1}. The cross-section formula for this kind of process is given as \cite{cross}
\begin{equation}
<\sigma v> = \frac{v^2 y^4 m_{\chi}^2}{48 \pi (m_{\chi}^2+m_{\psi}^2)^2},
\label{eq:14}
\end{equation}

\noindent where $y$ is Yukawa coupling of the interaction between DM and fermions, $m_\psi$ and $m_{\chi}$ represent the mass of Majorana fermion and relic particle mass respectively. Here, $ v$ is the relative velocity of two relic particles and is taken to be 0.3c at freeze out temperature. In case of m$_{DM}<M_W$, which indicates the low mass scale of relic particle, $\eta_2$,$\eta_3$ self annihilates via SM Higgs into the SM particles as shown in Fig.\ref{fig2}. The self annihilation cross-section is thus given as below\cite{relic2,cross1}\\
\begin{equation}
\sigma_{xx} = (\frac{|Y_f|^2 |\lambda_x|^2}{16 \pi s})(\frac{(s-4m_{f}^2)^{3/2}}{(s-4m_{x}^2)^{1/2}((s-4m_{h}^2)^2 + m_{h}^2 \Gamma_h^2)}),
\label{eq:15}
\end{equation}

\noindent where $ Y_f$ is Yukawa coupling of fermions and we have used its recent value as 0.308\cite{das}. Here, $ \Gamma_h$ is the SM Higgs decay width and its value used is 4.15 MeV. The $m_h$ is Higgs mass, that is 126 GeV, and $ x$ in Eqn.(\ref{eq:15}) represents $\eta_2,\eta_3$ and coupling of $ x$ with SM Higgs is represented as $\lambda_x$. Here, $s$ is thermally averaged center of mass squared energy and is given as \cite{cross1}\\
\begin{equation}
s = 4m_{\chi}^2 + m_{\chi}^2v^2.\\
\label{eq:16}
\end{equation}

\begin{figure}[t]
 %\hspace{-.5cm}
 \begin{subfigure}[b]{0.4\textwidth}
 \includegraphics[scale=.6]{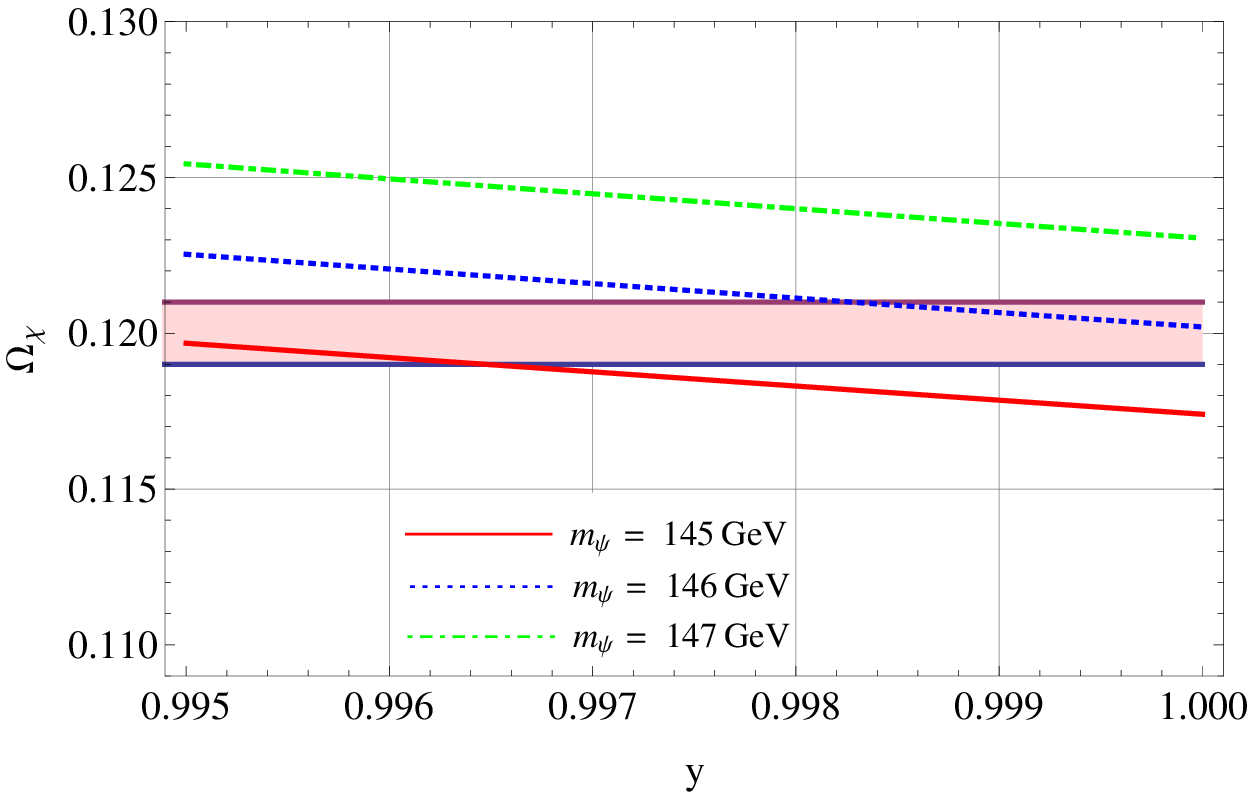}
 \caption{}
 \end{subfigure}
 ~\qquad
 \hspace{.5cm}
 \begin{subfigure}[b]{0.4\textwidth}
 \includegraphics[scale=.6]{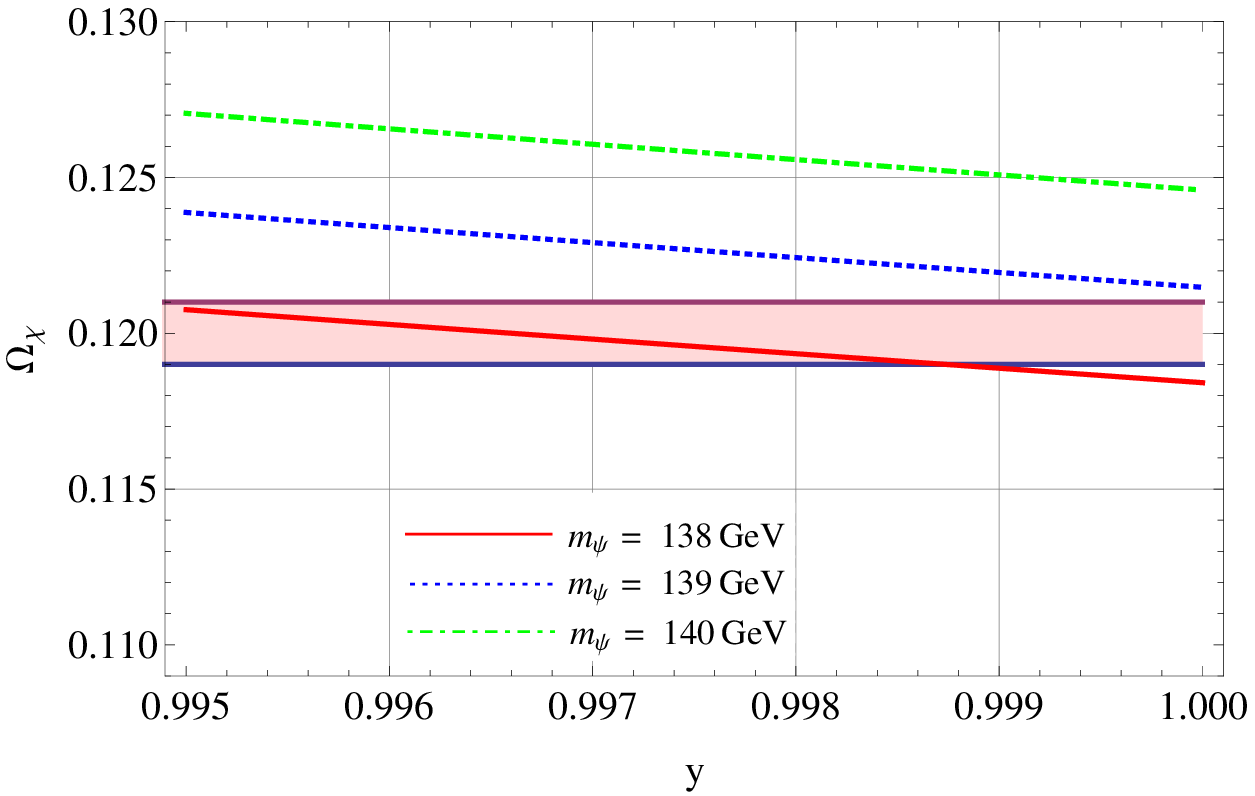}
 \caption{}
 \end{subfigure}
 
 \vspace{1cm}
 \begin{center}
\hspace{-2.5cm}
 \begin{subfigure}[b]{0.4\textwidth}
 \includegraphics[scale=.6]{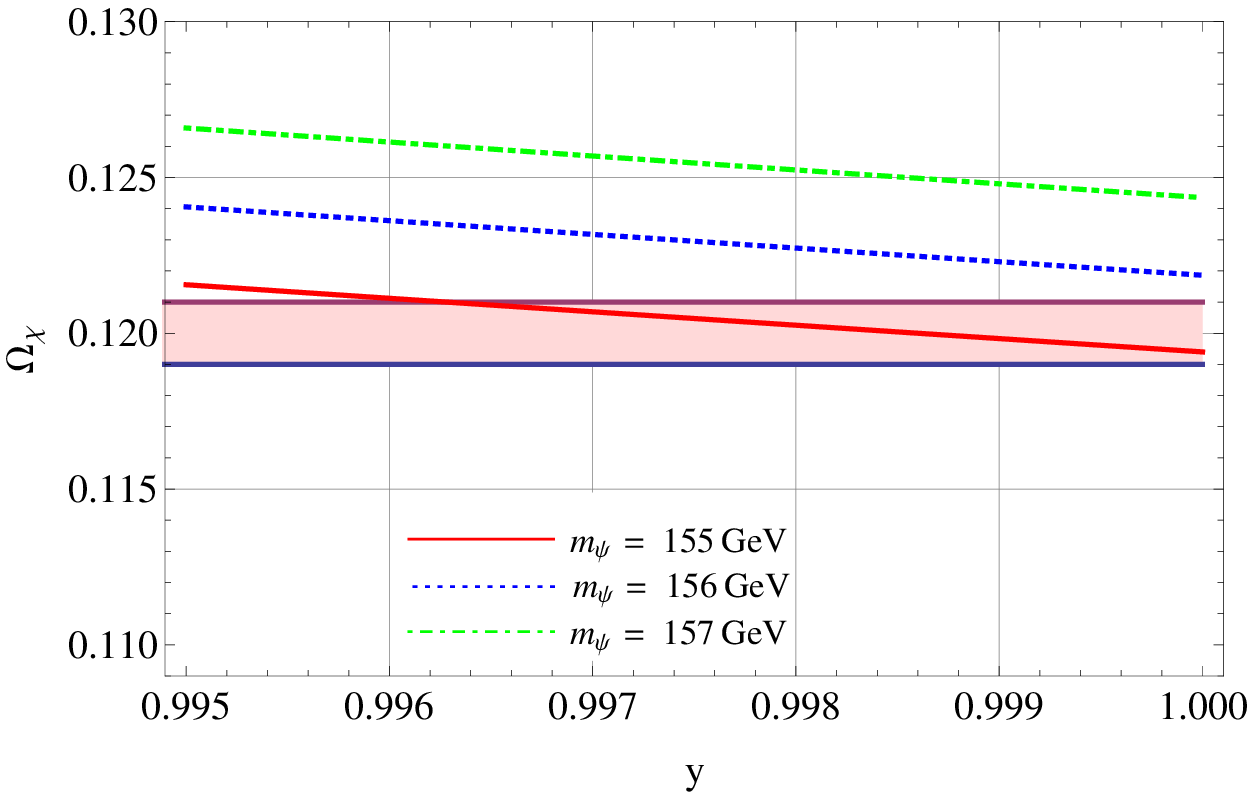}
 \caption{}
 \end{subfigure}
 \end{center}
 \caption{Relic density of DM vs Yukawa coupling plots for (a)DM mass(m$_{\chi})$ = 50 GeV, (b) DM mass(m$_{\chi})$ = 45 GeV and (c) DM mass(m$_{\chi})$ = 55 GeV.}
 \label{fig3}
 \end{figure}

\noindent The neutral component of scalar triplet $ \eta$ is our DM candidate as considered in \cite{52,53}. In this work we fixed our parameters in Eqn.(\ref{eq:13}) to obtain the recently updated constraints on relic abundance as reported by PLANCK 2018 data. To obtain the correct relic density of DM, we need to constrain the parameters like, Yukawa coupling, relic mass and mediator mass(right-handed neutrinos in our case). As stated above, we chose the relic mass much less than the mass of W-Boson. We have done our analysis for different values of relic mass and obtained different mediator masses and Yukawa couplings, which are shown in Fig.\ref{fig3}. This type of studies have already been done in \cite{cross,55}. In order to get correct relic abundance, we did our analysis for DM particle mass around 50 GeV, as suggested by many experimets like XENON1T\cite{Xenon}, PandaX-11\cite{Panda}, LUX\cite{Lux}, SuperCDMS\cite{CDMS} etc. For DM particle mass 45 GeV, 50 GeV and 55 GeV, we obtained the mass of right-handed neutrinos ranging from 138 GeV to 155 GeV. Yukawa coupling is obtained in the range 0.995-1. The results are shown in Table \ref{table3}.
 
 \begin{table}[t]
 \begin{center}
\begin{tabular}{|c|c|c|c|}
 \hline 
 S.No. & Relic Mass(m$_{\chi}$) & Mediator Mass(m$_{\psi}$) & Yukawa Coupling($y$) \\ 
 \hline 
 1 & 45 GeV & 138 GeV &  0.995-1\\ 
 \hline 
 2 & 50 GeV & 146 GeV & 0.998-1 \\ 
 \hline
 3 & 55 GeV & 155 GeV & 0.996-1\\
 \hline
 \end{tabular}  
 \caption{Constraints on relic(DM) mass, right-handed neutrino mass and Yukawa Coupling.}
 \label{table3}
 \end{center}
 \end{table}
 
 \section{Neutrinoless Double Beta($0\nu\beta\beta$) Decay }\label{sec:4}
 In general, the complex symmetric low energy effective neutrino mass matrix is given by
   \begin{equation}
   M_\nu=
   \begin{pmatrix}
M_{ee} & M_{e\mu} &  M_{e\tau} \\
M_{e\mu} & M_{\mu\mu} &  M_{\mu\tau} \\
M_{e\tau} &M_{\mu\tau}&  M_{\tau\tau} \\
 \end{pmatrix}.
 \label{eq:17}
   \end{equation}
 On comparing Eqn.(\ref{eq:17}) with the mass matrix in Eqn.(\ref{eq:11}) the effective Majorana neutrino mass appearing in $0\nu\beta\beta$ decay is
 \begin{equation}
     M_{ee}=X+X^{'},
     \label{eq:18}
 \end{equation}
 and
  \begin{equation}
     M_{\mu\tau}=X^{'},
     \label{eq:19}
 \end{equation}
  \begin{equation}
     M_{e\mu}=0,
     \label{eq:20}
 \end{equation}
 where $X$ and $X^{'}$ are inverse and type-II seesaw contributions to $0\nu\beta\beta$ decay. Eqn.(\ref{eq:20}) is our constraining condition to ascertain the allowed parameter space of the model. The elements of the mass matrix in Eqn.(\ref{eq:17}) are functions of low energy variables such as neutrino mass and mixing parameters and $CP$ violating phases. Using available data (Table \ref{table4}) for the known parameters and freely varying the unknown phases, we calculate $|M_{ee}|$ employing the formalism discussed below. Also, knowing $M_{\mu\tau}$ using data given in Table \ref{table4}, type-II contribution ($X^{'}$) can, independently, be calculated with the help of Eqn.(\ref{eq:19}). In this way, using Eqns.(\ref{eq:18}) and (\ref{eq:19}) we have been able to find the inverse ($X$) and type-II seesaw ($X^{'}$) contributions to $0\nu\beta\beta$ decay amplitude $|M_{ee}|$. 
 
\noindent In the charged lepton basis, the Majorana neutrino mass matrix can be written as 
\begin{equation}
M_{\nu}=UM_{d}U^T,
\end{equation}
where $M_{d}$ is diagonal mass matrix containing mass eigenvalues of neutrinos\\
$diag(m_{1}, m_{2}, m_{3})$. 
$ U$ is neutrino mixing matrix defined as $U=V.P$ 
where $ P$ is diagonal phase matrix  $diag(1,e^{i\alpha},e^{i(\beta+\delta)})$. In PDG representation, $V$ is given by
  \begin{equation}
   \begin{pmatrix}
c_{12} c_{13} & s_{12} c_{13} &  s_{13} e^{-i\delta} \\
-s_{12} c_{23} - c_{12} s_{23} s_{13} e^{i\delta} & c_{12} c_{23} - s_{12} s_{23} s_{13} e^{i\delta} &  s_{23} c_{13} \\
s_{12} s_{23} - c_{12} c_{23} s_{13} e^{i\delta} & -c_{12} s_{23} -s_{12} c_{23} s_{13} e^{i\delta} &  c_{23} c_{13} \\
 \end{pmatrix},
   \end{equation}
where $\delta$ is Dirac $CP$ violating phase and $\alpha$, $\beta$ are Majorana type $CP$ violating phases. The constraining condition (Eqn.(\ref{eq:20})) can be written as

\begin{equation}
c_{13} (e^{i(2 \beta + \delta)} m_3 s_{13} s_{23} - 
   c_{12} m_1 (c_{23} s_{12} + c_{12} e^{i\delta} s_{13} s_{23}) + 
   e^{2i\alpha} m_2 s_{12} (c_{12} c_{23} - e^{i\delta} s_{12} s_{13} s_{23}))=0.
\end{equation}
The above complex constraining equation gives two real constraints which can be solved for ratios $m_{2}/m_{1} \equiv R_{21}$ and $m_{3}/m_{1} \equiv R_{31}$ as
\begin{equation}
R_{21}=\frac{c_{12} (c_{12} s_{13} s_{23} \sin2 \beta+ c_{23} s_{12} \sin(2 \beta + \delta))}{s_{12} (s_{12} s_{13} s_{23} \sin2 (\alpha - \beta) - 
   c_{12} c_{23} \sin(2 \alpha - 2 \beta - \delta))},
\end{equation}
and 
\begin{equation}
    R_{31}=\frac{c_{12} (c_{12} s_{12} (-c_{23}^2 + s_{13}^2 s_{23}^2) \sin2 \alpha + 
   c_{23} s_{13} s_{23} (-c_{12}^2 \sin(2 \alpha - \delta) + 
      s_{12}^2 \sin(2 \alpha + \delta)))}{s_{13} s_{23} (s_{12} s_{13} s_{23} \sin2 (\alpha - \beta) - 
   c_{12} c_{23} \sin(2 \alpha - 2 \beta - \delta))}.
\end{equation}
 Using these mass ratios, we have two different values of lowest eigenvalue $m_1$, equating them gives us 
\begin{equation}
R_{\nu}\equiv\frac{\Delta m_{21}^2}{|\Delta m_{32}^2|}=\frac{R_{21}^2-1}{|R_{31}^2+R_{21}^2-2|}.
\end{equation}
The parameter space is constrained using the 3$\sigma$ range of $ R_{\nu}$. Neutrino oscillation parameters ($\theta_{12}$, $\theta_{23}$, $\theta_{13}$, $\Delta m_{21}^2$, $\Delta m_{32}^2$) are randomly generated with Gaussian distribution while $CP$ phases ($\delta$,$\alpha$,$\beta$) are varied randomly in their full range with uniform distribution. \\
The neutrino masses can be obtained using mass-squared differences as\\

\begin{center}
     $ m_2=\sqrt{m_1^{2}+\Delta m_{21}^{2}}$,\hspace{.4cm}$m_3=\sqrt{m_1^{2}+\Delta m_{31}^{2}}$ \hspace{0.3cm} for Normal ordering (NO),\\
   \end{center}
  and
    \begin{center}
   $m_1=\sqrt{m_{3}^{2}-\Delta m_{31}^{2}}$ ,\hspace{.4cm} $m_2=\sqrt{m_3^{2}-\Delta m_{31}^{2}+\Delta m_{21}^{2}}$ \hspace{0.3cm} for Inverted ordering (IO),
\end{center}
whereas the lightest neutrino mass ($m_1$(NO), $m_3$(IO)) is obtained from mass ratios.
Also, we calculate the effective mass parameter
\begin{equation}
|M_{ee}|=\left|m_{1} c_{12}^2 c_{13}^2 + m_{2} s_{12}^2 c_{13}^2 e^{2i \alpha}+ m_{3} s_{13}^2 e^{2i \beta}\right|,
\end{equation}
and $|M_{\mu\tau}|$ for the allowed parameter space which are, further, used to find inverse ($|X|$) and type-II ($|X^{'}|$) seesaw contribution to $|M_{ee}|$.

\begin{table}[t]
\begin{center}
\begin{tabular}{c|c|c}
\hline \hline 
Parameter & Best fit $\pm$ \( 1 \sigma \) range & \( 3 \sigma \) range  \\
\hline \multicolumn{2}{c} { Normal neutrino mass ordering \( \left(m_{1}<m_{2}<m_{3}\right) \)} \\
\hline \( \sin ^{2} \theta_{12} \) & $0.304^{+0.013}_{-0.012}$ & \( 0.269-0.343 \)  \\
\( \sin ^{2} \theta_{13} \) & $0.02221^{+0.00068}_{-0.00062}$ & \( 0.02034-0.02420 \) \\
\( \sin ^{2} \theta_{23} \) & $0.570^{+0.018}_{-0.024}$ & \( 0.407-0.618 \)  \\
\( \Delta m_{21}^{2}\left[10^{-5} \mathrm{eV}^{2}\right] \) & $7.42^{+0.21}_{-0.20}$& \( 6.82-8.04 \) \\
\( \Delta m_{31}^{2}\left[10^{-3} \mathrm{eV}^{2}\right] \) & $+2.541^{+0.028}_{-0.027}$ & \( +2.431-+2.598 \) \\
\hline \multicolumn{2}{c} { Inverted neutrino mass ordering \( \left(m_{3}<m_{1}<m_{2}\right) \)} \\
\hline \( \sin ^{2} \theta_{12} \) & $0.304^{+0.013}_{-0.012}$ & \( 0.269-0.343 \)\\
\( \sin ^{2} \theta_{13} \) & $0.02240^{+0.00062}_{-0.00062}$ & \( 0.02053-0.02436 \) \\
\( \sin ^{2} \theta_{23} \) & $0.575^{+0.017}_{-0.021}$& \( 0.411-0.621 \) \\
\( \Delta m_{21}^{2}\left[10^{-5} \mathrm{eV}^{2}\right] \) & $7.42^{+0.21}_{-0.20}$ & \( 6.82-8.04 \) \\
\( \Delta m_{32}^{2}\left[10^{-3} \mathrm{eV}^{2}\right] \) & $-2.497^{+0.028}_{-0.028}$ & \( -2.583--2.412 \)  \\
\hline \hline
\end{tabular}
\end{center}
\caption{Neutrino oscillations experimental data NuFIT 5.0 used in the numerical analysis\cite{data}.}
\label{table4}
\end{table}
\begin{figure}[t]
	\begin{center}
			{\epsfig{file=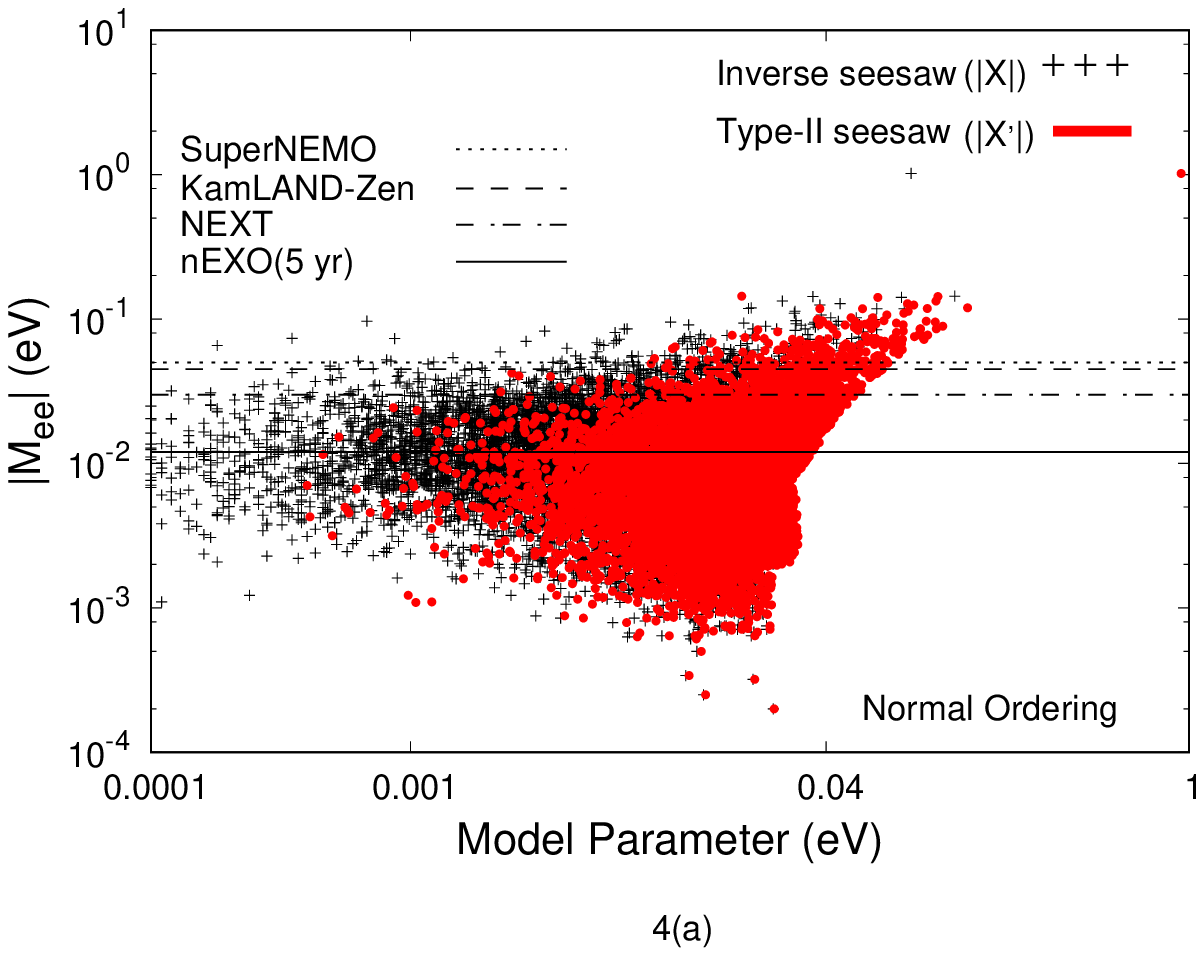,height=7.0cm,width=7.0cm}
				\epsfig{file=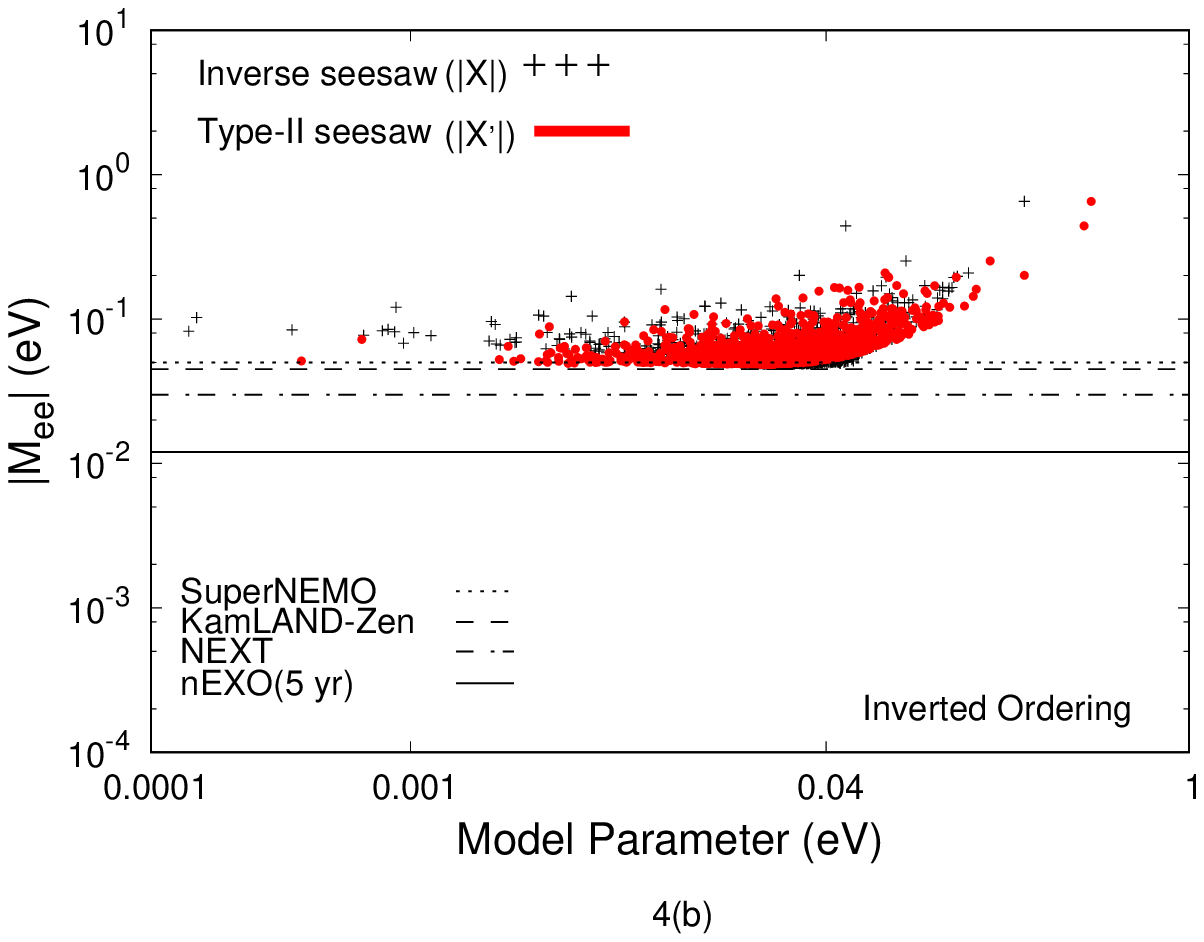,height=7.0cm,width=7.0cm}}
		\end{center}
\caption{\label{fig:4} Variation of effective Majorana neutrino mass $|M_{ee}|$ with Contribution of inverse seesaw and type-II seesaw model parameters ($|X|$ and $|X'|$), for both normal and inverted ordering.}
\end{figure}

\noindent In Fig.\ref{fig:4}, we have plotted effective mass parameter $|M_{ee}|$ with inverse seesaw contribution (X) and type-II seesaw contribution (X$'$). The sensitivity reach of 0$\nu\beta\beta$ decay experiments like SuperNEMO \cite{nemo}, KamLAND-Zen \cite{zen}, NEXT \cite{next1,next2},
	nEXO\cite{nexo} is, also, shown in Fig.\ref{fig:4}. In Fig.\ref{fig:4}(a), it is clear that for NO, higher density of points for type-II seesaw indicate contribution  of $\mathcal{O}(0.01eV)$, whereas inverse seesaw contribution is of $\mathcal{O}(0.001eV)$. On the other hand, for IO, different seesaw contributions are of same order as can be seen in Fig.\ref{fig:4}(b). Hence, type-II seesaw contribution to $|M_{ee}|$ is large as compared to inverse seesaw contribution for NO for the texture one-zero model within the framework of inverse seesaw and type-II seesaw. For normal ordering neutrino mass spectrum, the $|M_{ee}|$ goes below upto the $\mathcal{O}(10^{-4}eV)$ and we did not obtain a clear lower bound as shown in Fig.\ref{fig:4}(a). On contrary, for inverted ordering, there is a clear cut lower bound for the $|M_{ee}|$ which can be probed in future $0\nu\beta\beta$ decay experiments. It can be seen from Fig.\ref{fig:4}(b), that the $0\nu\beta\beta$ decay experiments like SuperNEMO, KamLAND-Zen can probe the inverted ordering spectrum.

	\begin{figure}[H]
	\begin{center}
			{\epsfig{file=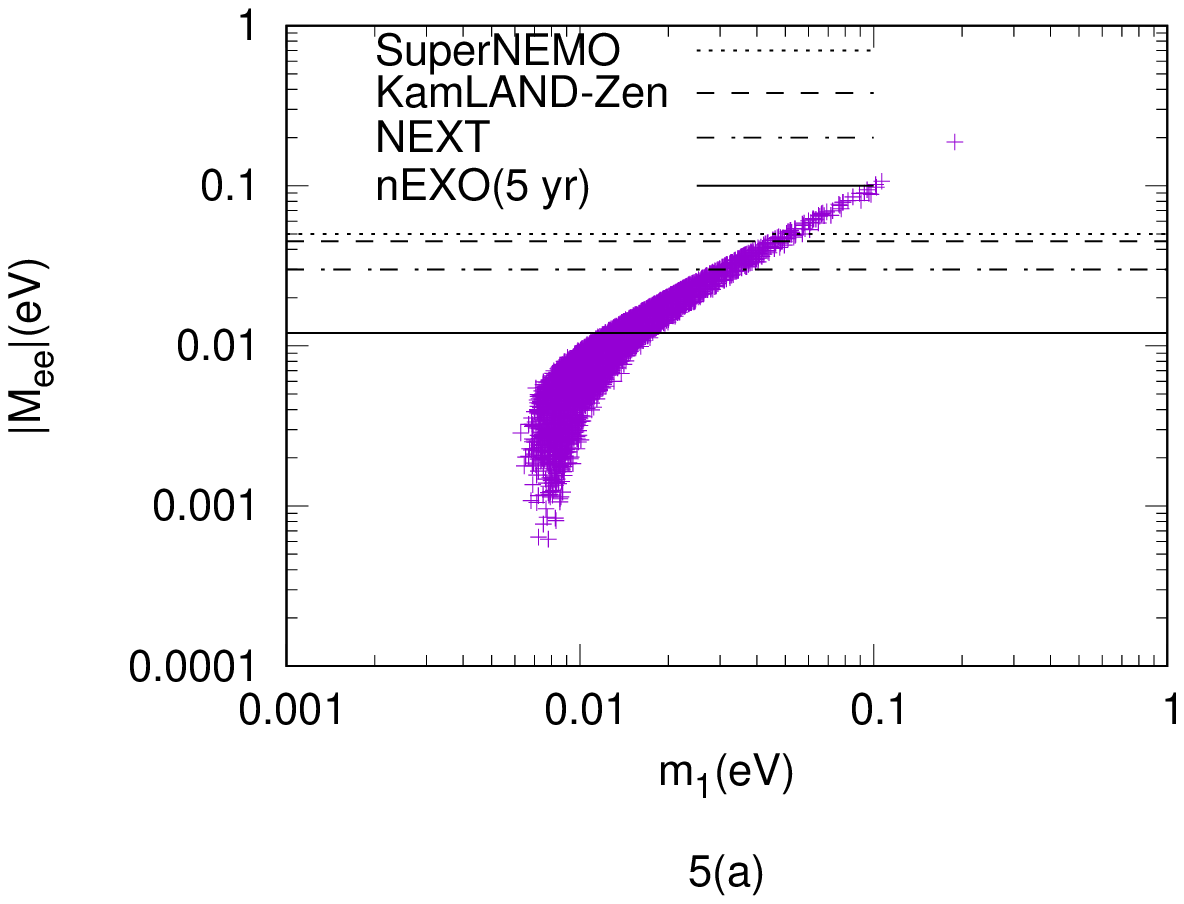,height=7.0cm,width=7.0cm}
				\epsfig{file=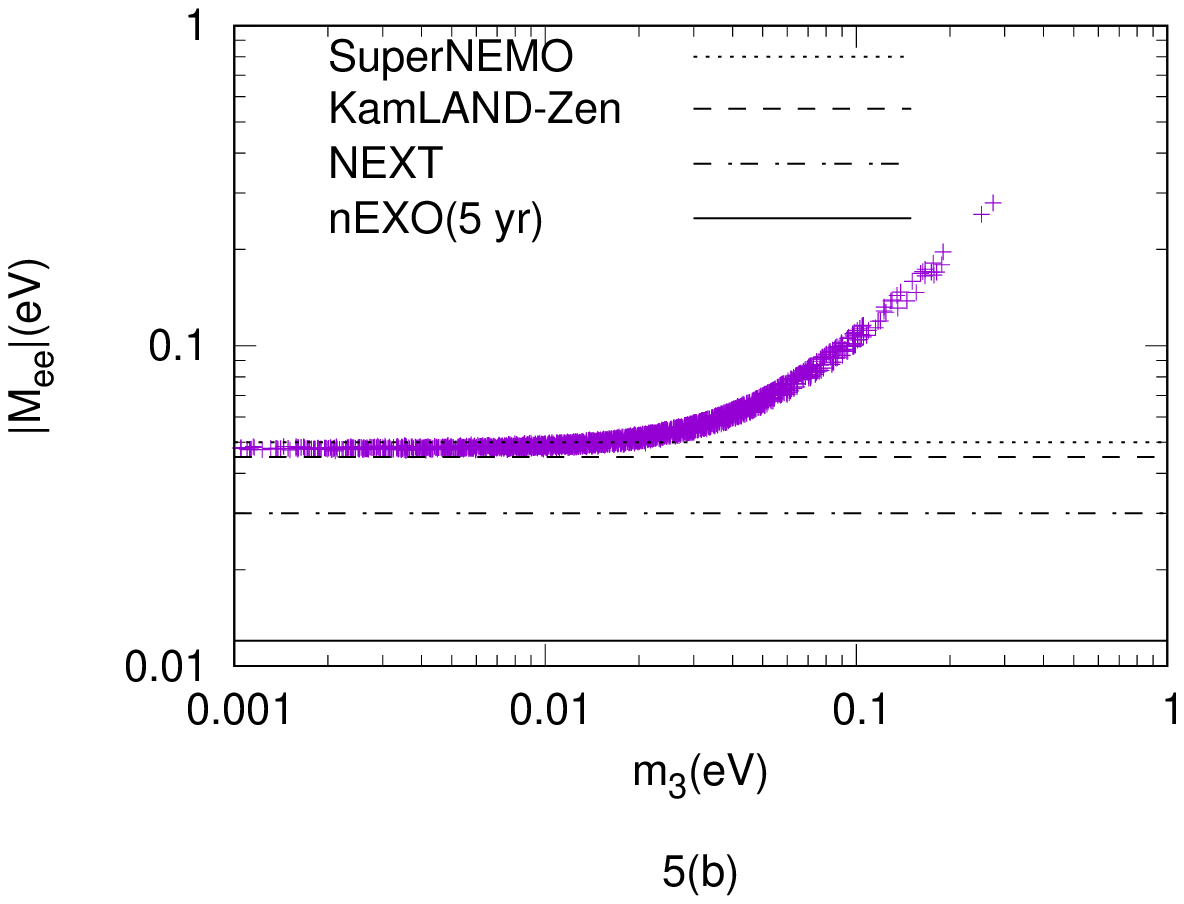,height=7.0cm,width=7.0cm}}
		\end{center}
\caption{\label{fig:5} Variation of effective Majorana neutrino mass $|M_{ee}|$ with lightest neutrino mass $m_1$($m_3$) for both normal(inverted) ordering of neutrino masses.}
\end{figure}

	\begin{figure}[H]
	\begin{center}
			{\epsfig{file=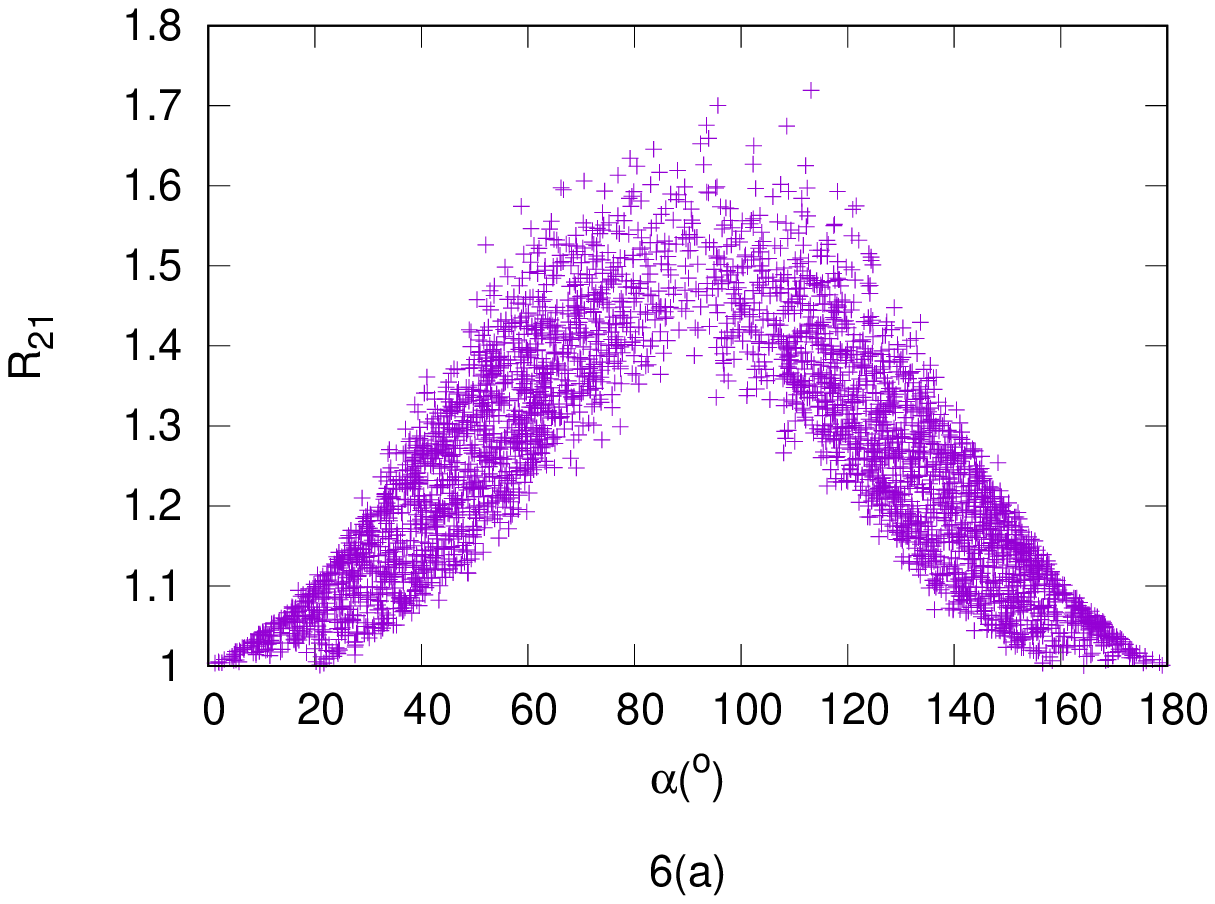,height=7.0cm,width=7.0cm}
				\epsfig{file=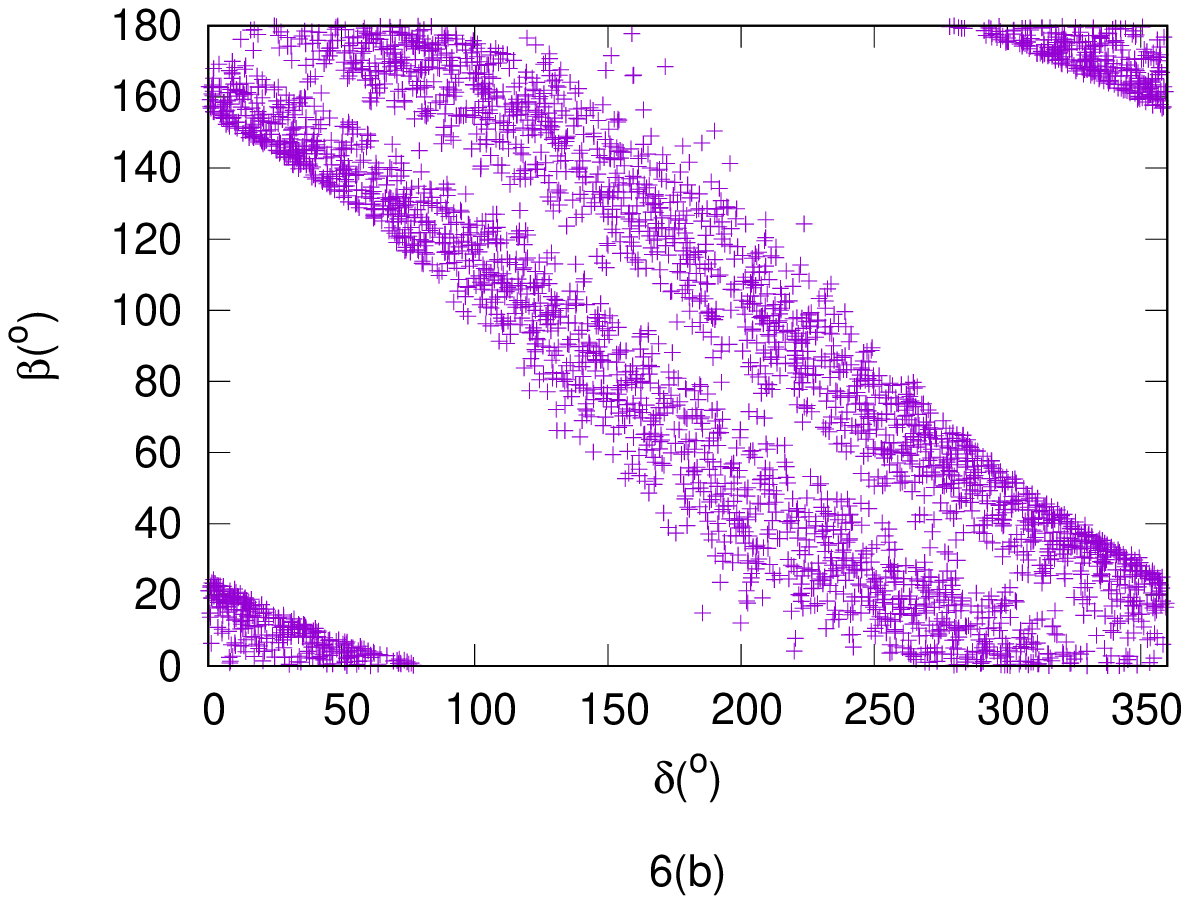,height=7.0cm,width=7.0cm}}
					{\epsfig{file=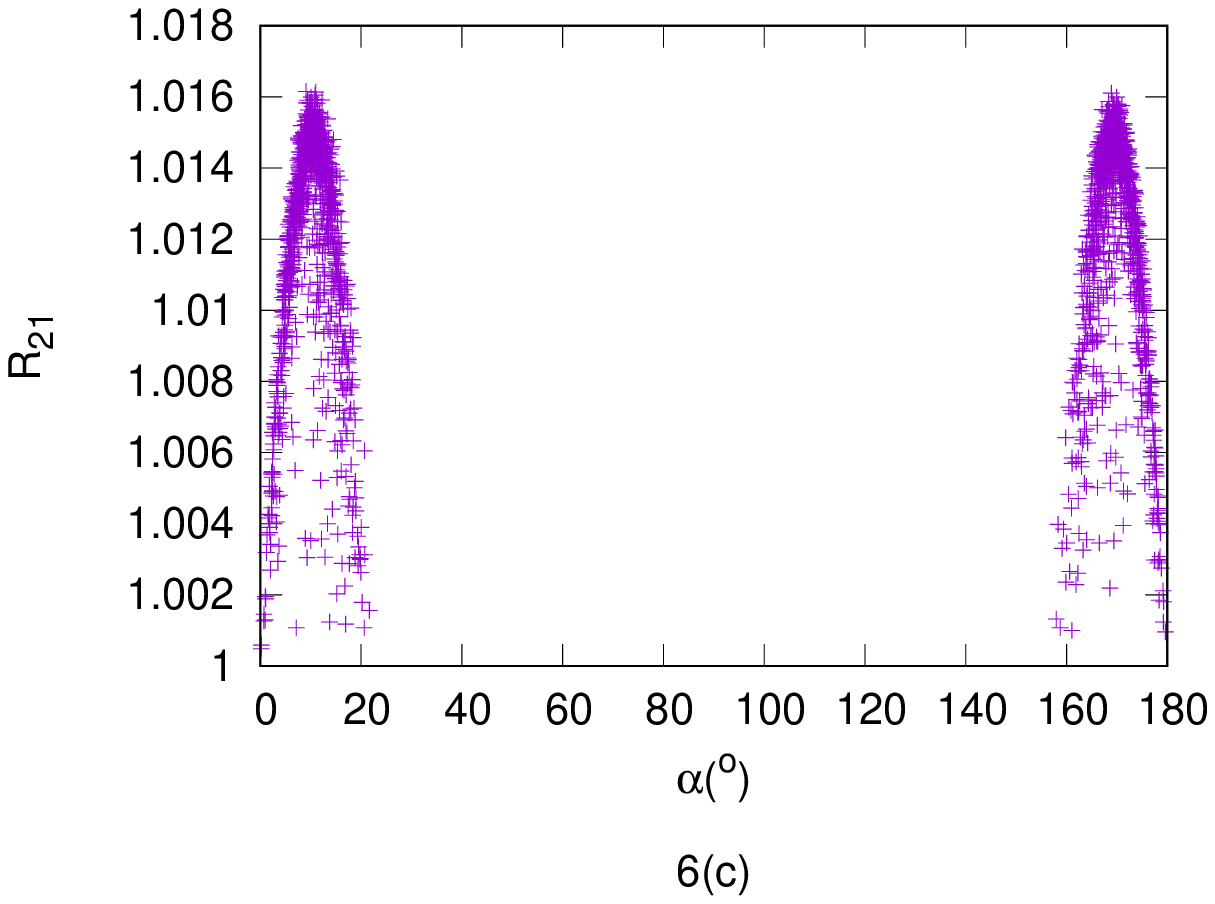,height=7.0cm,width=7.5cm}
				\epsfig{file=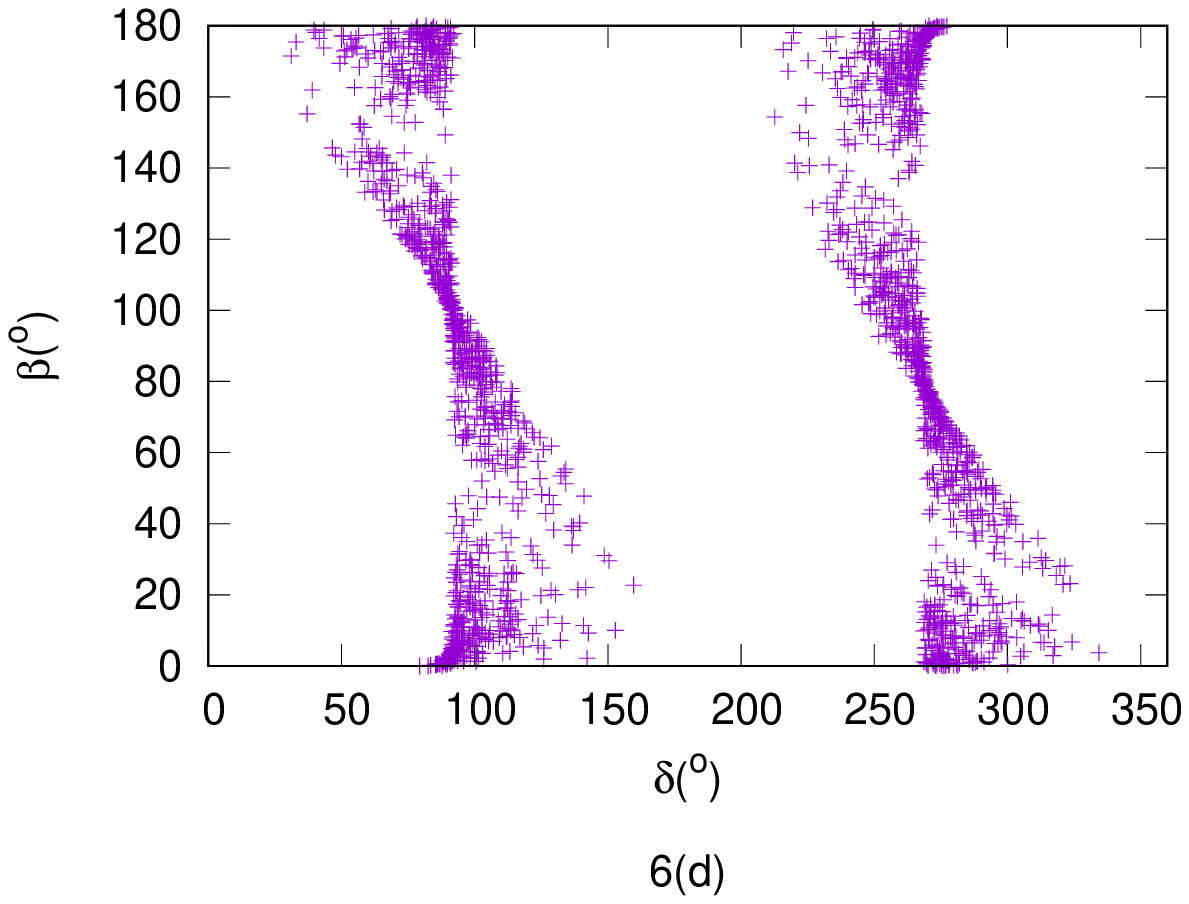,height=7.0cm,width=7.0cm}}
		\end{center}
\caption{\label{fig:6} Correlation plots of ($\alpha$-$R_{21}$) and ($\delta$-$\beta$). First(second) row depicts the correlations for normal(inverted) ordering of neutrino masses.}
\end{figure}

	 \noindent In Fig.\ref{fig:5}, we have depicted the correlation of effective Majorana mass parameter $|M_{ee}|$ with the lightest neutrino mass $m_1 (m_3)$ for normal (inverted) ordering of neutrino masses. $|M_{ee}|$ goes below up to 0.001eV for normal ordering whereas there exist a lower bound near 0.04eV for inverted ordering. The sensitivity reaches of various current and future $0\nu\beta\beta$ decay experiments have, also, been shown in Fig.\ref{fig:5}. Although, for normal ordering, $|M_{ee}|$ is beyond the sensitivity reach of experiments but non-observation $0\nu\beta\beta$ decay shall rule out the inverted ordering in this model.
	 
\begin{figure}[t]
	\begin{center}
			{\epsfig{file=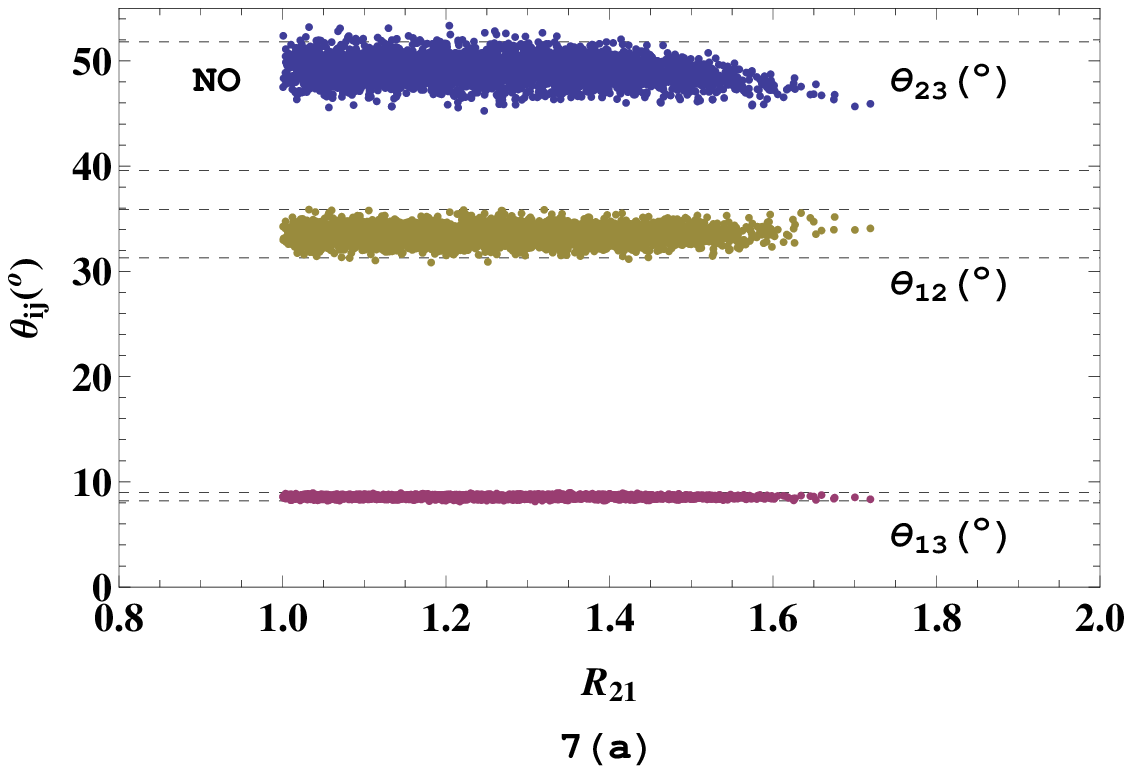,height=7.0cm,width=7.0cm}
				\epsfig{file=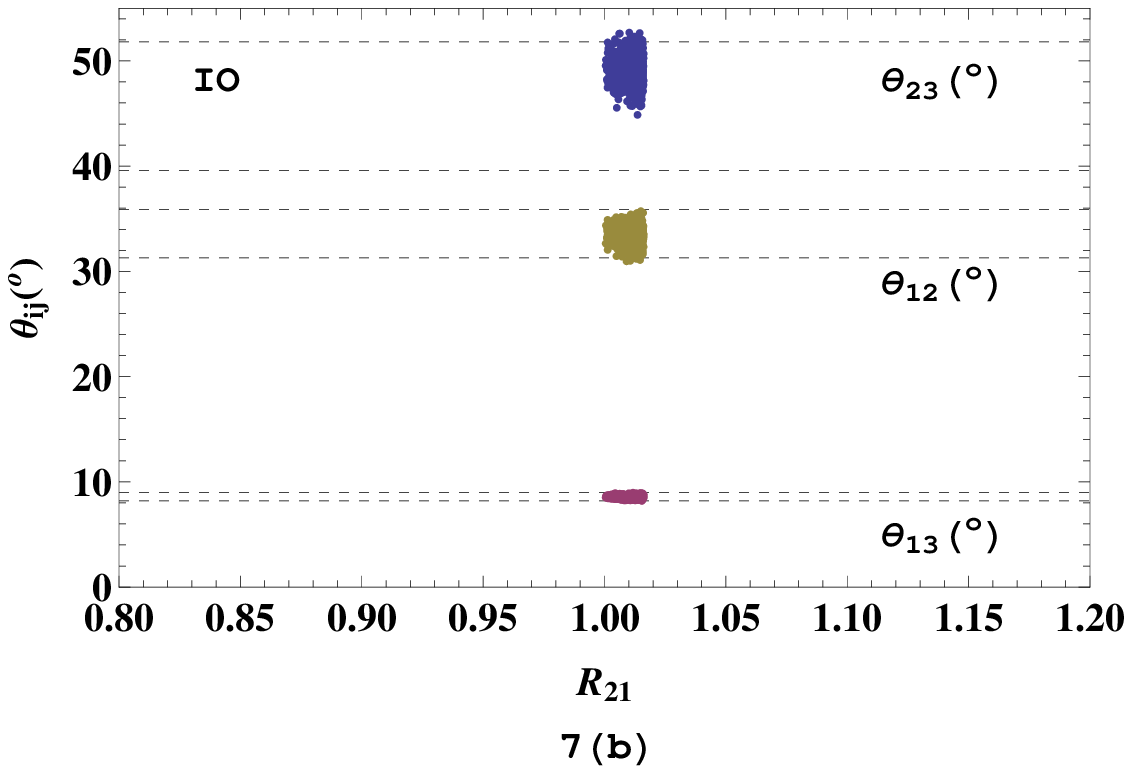,height=7.0cm,width=7.0cm}}
		\end{center}
\caption{\label{fig:7} Correlation plots of ($\theta_{ij}$-$R_{21}$) for normal(inverted) ordering of neutrino masses. Horizontal dashed lines are $3\sigma$ experimental range of respective mixing angles. }
\end{figure}

	 \noindent Furthermore, we have studied the correlations amongst the Majorana phases($\alpha,\beta$), Dirac phase($\delta$) and mass ratio($R_{21}\equiv\frac{m_2}{m_1}$). Some representative plots are given in Fig.\ref{fig:6}. The correlation plots ($\alpha-R_{21}$) and ($\delta-\beta$) are shown in Fig.\ref{fig:6}(a)(\ref{fig:6}(c)) and Fig.\ref{fig:6}(b)(\ref{fig:6}(d)), respectively for normal(inverted) mass ordering. For normal ordering, there is no preferable range of CP-phase(Dirac or Majorana) whereas, in case of inverted ordering, Majorana CP-phase $\alpha$ is constrained in the range($0^{\circ}-20^{\circ}) \cup (160^{\circ}-180^{\circ})$ and $\delta$ is found to lie in $(40^{\circ}-150^{\circ})\cup (200^{\circ}-325^{\circ}$) range. In Fig.\ref{fig:7}, we have depicted the correlation of neutrino mixing angles with mass ratio $R_{21}$ for normal(inverted) ordering. The model satisfies the neutrino oscillation data for both mass ordering. The model predictions are consistent with the model independent analysis performed in Ref.\cite{text015}.

 \section{Conclusions}\label{sec:5}
In conclusion, we have proposed a model based on $A_4$ discrete flavor symmetry implementing inverse and type-II seesaw mechanisms to have LHC accessible TeV scale right-handed neutrino mass and texture one-zero in the resulting Majorana neutrino mass matrix, respectively. The symmetry is broken down to $Z_2$ subgroup i.e. $G_f\equiv A_4\times Z_{2}^{'}\times Z_3\times Z_4\rightarrow Z_2$  by the $vev$ $\langle \eta \rangle \sim v_{\eta} (1,0,0)$ which, further stabilizes the DM candidate $\eta_{2,3}$. We scanned the ranges of Yukawa coupling ($y$), right-handed neutrino mass ($m_\psi$) and DM mass ($m_{\chi}$) to obtain observed relic abundance of DM $\eta_{2,3}$.  We find that to have observed relic abundance the $m_\chi$, $m_\psi$ and $y$ should be around 50 GeV, 146 GeV and 0.998-1, respectively. Also, we have obtained the prediction of the model for $0\nu\beta\beta$ decay amplitude $|M_{ee}|$. In particular, we calculated inverse ($X$) and type-II ($X^{'}$) seesaw contributions to $|M_{ee}|$, while texture one-zero model being consistent with low energy experimental constraints. The type-II seesaw contribution to $|M_{ee}|$ is found to be large as compared to inverse seesaw contribution for normal ordering neutrino masses. The model predicts a robust lower bound on $|M_{ee}|$ for inverted ordering  neutrino masses which can be probed in future $0\nu\beta\beta$ decay experiments like SuperNEMO, KamLAND-Zen. The correlation of $|M_{ee}|$-$m_1(m_3)$ shows that inverted ordering is ruled out in case of non-observation of $0\nu\beta\beta$ decay while normal ordering may still be allowed. Also, contradistinction to normal ordering, the CP-violating phases $\delta$ and $\alpha$ are constrained in the range ($0^{\circ}-20^{\circ}) \cup (160^{\circ}-180^{\circ})$ and $(40^{\circ}-150^{\circ})\cup (200^{\circ}-325^{\circ}$), respectively, for inverted ordering of neutrino masses.\\
 
 \hspace{-.4cm}\textbf{\Large{Acknowledgments}}
 \vspace{.3cm}\\
 R. Verma acknowledges the financial support provided by the Central University of Himachal Pradesh. B. C. Chauhan is thankful to the Inter University Centre for Astronomy and Astrophysics (IUCAA) for providing necessary facilities during the completion of this work. M. K. acknowledges the financial support provided by Department of Science and Technology, Government of India vide Grant No. DST/INSPIRE Fellowship/2018/IF180327.

\end{document}